\documentclass[final,5p,times,twocolumn,numbers]{elsarticle}

\usepackage{amssymb}
\usepackage{lipsum}
\usepackage{url}
\usepackage{here}
\usepackage{multicol}
\usepackage[colorlinks=true,citecolor=black,urlcolor=blue]{hyperref}
\usepackage{subcaption}

\usepackage{lineno}
\usepackage{fouriernc}

\journal{Nuclear Physics A}

\begin{document}

\begin{frontmatter}



\title{The energy dependence of cluster size and its physical processes \\in the proton measurement with TimePix3 silicon detector}


\affiliation[first]{organization={Department of Physics and Astronomy, Tokyo University of Science},
            addressline={2641 Yamazaki}, 
            city={Noda},
            postcode={278-8510}, 
            state={Chiba},
            country={Japan}}

\affiliation[IPMU]{organization={Kavli Institute for the Physics and Mathematics of the Universe (Kavli IPMU, WPI), The University of Tokyo},
            addressline={5-1-5 Kashiwanoha}, 
            city={Kashiwa},
            postcode={277-8583}, 
            state={Chiba},
            country={Japan}}

\affiliation[RIKEN]{organization={RIKEN},
            addressline={Hirosawa 2-1}, 
            city={Wako},
            postcode={351-0198}, 
            state={Saitama},
            country={Japan}}

\affiliation[UCB]{organization={Space Sciences Laboratory, University of California, Berkeley},
            addressline={7 Gauss Way}, 
            city={Berkeley},
            postcode={94720}, 
            state={CA},
            country={USA}}
            
\affiliation[UT]{organization={Department of Physics, The University of Tokyo},
            addressline={7-3-1 Hongo}, 
            city={Bunkyo},
            postcode={113-0033}, 
            state={Tokyo},
            country={Japan}}

\affiliation[chubu]{organization={Chubu University},
            city={Kasugai},
            postcode={487-8501}, 
            state={Aichi},
            country={Japan}}

\affiliation[nifs]{organization = {National Institute for Fusion Science (NIFS)},
            city = {Toki},
            postcode = {509-5292},
            state = {Gifu},
            country = {Japan}}

\affiliation[kyushu]{organization={Department of Physics, Kyushu University},
            addressline={744 Motooka, Nishi-Ku}, 
            city={Fukuoka},
            postcode={819-0395}, 
            state={Fukuoka},
            country={Japan}}

\author[first]{Naoki Itoh}
\author[IPMU]{Hugo Allaire}
\author[RIKEN]{Tokihiro Ikeda}
\author[UCB,UT,IPMU]{Shunsaku Nagasawa}
\author[chubu,nifs]{Shinji Okada}
\author[IPMU]{Tadayuki Takahashi}
\author[RIKEN,kyushu]{Aiko~Takamine}
\author[chubu]{Yuichi Toyama}
\author[first]{Yuusuke Uchida}
\author[RIKEN]{Hideki Ueno}

\begin{abstract}
We investigated the energy dependence of the number of triggered pixels, or cluster size, when charged particles are detected using the TimePix3 detector with a silicon sensor.
We measured protons in the range of 1.5–3.3 $\mathrm{MeV}$ from a Pelletron accelerator at RIKEN using a TimePix3 detector with a 500 $\mathrm{\othermu m}$-thick silicon sensor.
We determined from the experimental results a cluster size comprised between 30 and 80 pixels.
To understand the physical process that produces large cluster images and its energy dependence, we simulated the charge carrier drifts in the sensor, assuming the incidence of a proton in the detector.
The cluster sizes estimated in the simulation were smaller than those observed in the experiment, and remained constant across the entire energy range, when thermal diffusion and charge carriers self-repulsion were considered as the factors of the cluster image formation.
In addition, we discovered that the size of the cluster image and its energy dependence observed in the experiment could be well explained when considering that the TimePix3 detector is sensitive to the transient induced charges, allowing even pixels that do not collect the charge carriers to trigger.
We conclude that the cluster size measurement is a promising method for evaluating the energy deposited by a charged particle in the TimePix3 detector.
\end{abstract}




\end{frontmatter}




\section{Introduction}\label{Section1}
\vspace{-5pt}
Pixelated silicon semiconductor detectors can measure charged particles and are used in various fields, including nuclear and particle physics experiments \cite{Review_TimePix, 3Dtrack, Measurementofthe212Po,LET}.
The TimePix3 detector \cite{65k} is one of these detectors, developed by the MediPix Collaboration3 \cite{MediPix3Collaboration}, CERN, Nikhef and the University of Bonn as a successor to the TimePix detector \cite{65k_timepix}.
A TimePix3 detector consists of a silicon sensor and a TimePix3 readout chip, divided into $256 \times 256$ pixels with a $55~\mathrm{\othermu m}$ pitch.
The detector collects holes to the pixel electrodes.
It employs ToT (Time over Threshold) and ToA (Time of Arrival) methods to determine the energy and arrival time of radiation. 
By using the data-driven mode of TimePix3, we can obtain the arrival time of the radiation with a 1.56 $\mathrm{ns}$ temporal resolution and the energy deposition in each pixel, simultaneously.\\
\indent
The ToT method provides a wide dynamic range for the energy measurements.
Therefore, this instrumental method can be used to evaluate the energy of various types of charged particles across extended energy ranges, from light ions to heavy ions, in addition to X-ray and $\gamma$-ray \cite{Review_TimePix}. 
However, there are two issues affecting high energy measurements using TimePix3 detectors.
The first one is the difficulty to calibrate ToT values corresponding to high energies.
The detection efficiency of a silicon sensor with a thickness of several hundred $\mathrm{\othermu m}$ is low for high-energy photons.
For example, the detection efficiency of a 500 $\mathrm{\othermu m}$ silicon sensor is less than 4 \% for 60 $\mathrm{keV}$ photons, according to calculations performed with the Geant4 toolkit \cite{geant4}.
In addition, charge clouds generated by the incidence of high-energy radiations spread across multiple pixels.
These physical effects make the calibration of ToT values corresponding to energies above 60 $\mathrm{keV}$ inefficient, as the calibration function of the TimePix3 detector cannot be expressed by a monotonically increasing linear function.
The second issue is that high energy deposition leads to a saturation or a decrease of ToT values \cite{volcano1,volcano2,volcano3}.
The ToT value is measured using a 10-bit clock signal for each pixel and the range of the ToT value is 0 to 1022 \cite{65k}.
If the energy deposited in a pixel is greater than the amount corresponding to a ToT value of 1022, the ToT value will saturate.
Moreover, if the energy deposited in a pixel exceeds 500 $\mathrm{keV}$, the ToT value will decrease.
This phenomenon is observed when measuring high-energy charged particles and is called ``the volcano effect``.
A pixel with a large amount of energy deposited produces a short-width pulse from the preamplifier, resulting in a low ToT value.
Owing to these two issues, it is impractical to employ ToT values to evaluate the energy of high-energy charged particles.\\
\indent
As an alternative criterion to the ToT measurement for energy evaluation, we consider using the cluster size, defined as the number of triggered pixels comprised in the image.
When charged particles are measured with a TimePix3 detector, they are observed as a cluster image that is spread over several pixels.
The cluster image is like a disc when the charged particles are vertically incident on the detector surface.
The cluster size varies depending on various physical parameters of the detector and of the charged particle; however, the primary driver is the energy deposited in the sensor \cite{Chargeshare1,Chargeshare2}.\\
\indent
In this paper, we report the results of a proton irradiation experiment to evaluate the energy spectral performance and the energy dependence of cluster size (Section \ref{Section2}).
Similarly, we further report the results of the cluster size estimation using the charge carrier drift simulation to understand the physical process behind the cluster image formation (Section \ref{Section3}).
In Section \ref{discuss}, we discuss the physical factors influencing the cluster image formation when protons are detected using the TimePix3 detector, based on the experimental and the simulation results.

\section{Proton Measurement}\label{Section2}
\subsection{Beam condition}
\begin{figure}[h!]
	\centering 
	\includegraphics[scale=0.8,angle=0]{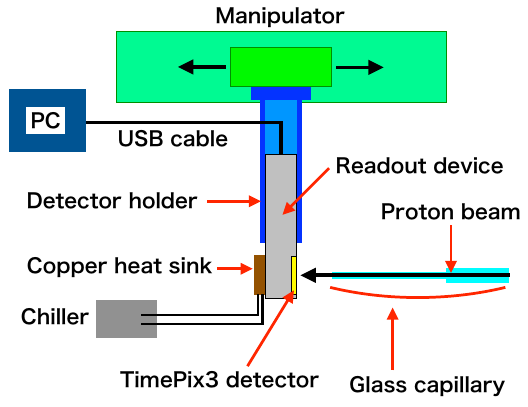}
	\caption{Experimental setup. All equipment is placed in the air. A vacuum is maintained in the RIKEN tandem accelerator and the glass capillary.}
	\label{Experiment setup}
\end{figure}

\begin{figure}[h!]
    \centering
    \begin{subfigure}{0.49\linewidth}
        \centering
        \includegraphics[width=\linewidth]{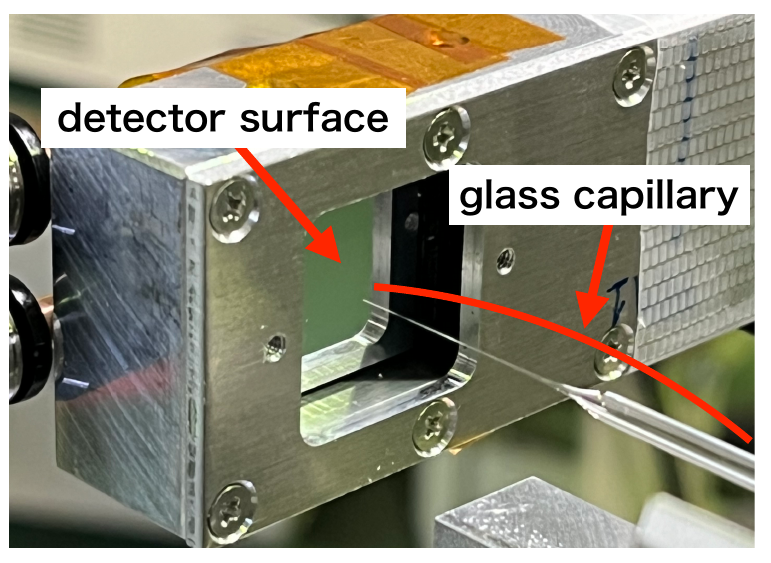}
        \caption{}
        \label{sensor_and_glass_capillary}
    \end{subfigure}
    \centering
    \begin{subfigure}{0.49\linewidth}
        \includegraphics[width=\linewidth]{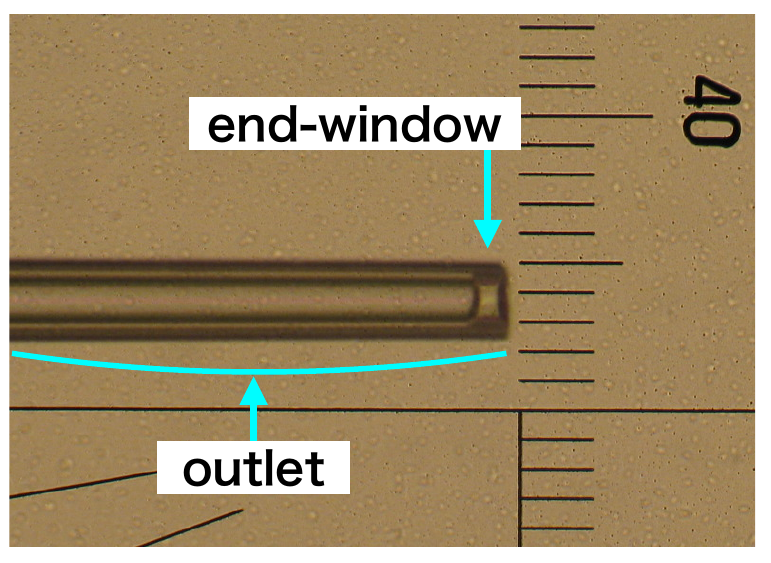}
        \caption{}
        \label{capillary}
    \end{subfigure}
    \caption{(a) Detector surface and glass capillary. (b) Outlet of the glass capillary. The inner diameter of the outlet and thickness of the end-window are 10 $\mathrm{\othermu m}$. The inner diameter appears larger than 10 $\mathrm{\othermu m}$ due to the refractive index of plastic \cite{capillary1,capillary2}.}
\end{figure}

We employed a proton beam line from the RIKEN tandem accelerator (Pelletron 5SDH-2, 1.7 $\mathrm{MV}$ max.).
Protons were irradiated onto the TimePix3 detector as a microbeam using a vacuumed glass capillary.\cite{capillary1,capillary2}.
Figure \ref{Experiment setup} shows the schematic diagram of the experimental setup.
The diameter of the capillary outlet is 10 $\mathrm{\othermu m}$, and it is sealed with a 10 $\mathrm{\othermu m}$ thick end-window (plastic film) to maintain a vacuum (Figure \ref{capillary}).
The energy value of protons from the accelerator was set to 3.4 $\mathrm{MeV}$; however, proton's energy after exiting the glass capillary was lower owing to energy losses occurring as the protons passed through the capillary end-window.
In addition, there is a 0.5 $\mathrm{\othermu m}$ aluminum film serving as a common electrode on the sensor, causing additional energy attenuation.\\
\indent
In this experiment, we prepared energy values of 3240, 2741, 2348, 1920, and 1513 $\mathrm{keV}$ by setting the distance from the detector surface to the capillary end-window \cite{capillary1, capillary2} at 4, 44, 72, 99, and 121 $\mathrm{mm}$ in the air, respectively.
We calculated the energy losses in the glass capillary end-window, aluminum film, and air using SRIM (The Stopping and Range of Ions in Matter) \cite{SRIM}.

\subsection{Experimental detector conditions}
During all measurements, the detector operated in the data-driven mode with a high voltage of 165 $\mathrm{V}$ to adequately form the depletion layer, and the trigger threshold level was set to 3.0 $\mathrm{keV}$.\\
\indent
We used a MiniPix TPX3 Flex from Advacam \cite{advacam} as a TimePix3 detector with integrated readout electronics.
Advacam provided us the calibration data for each individual pixel and we used this data to convert the ToT value to the energy.
This calibration data was confirmed to be valid up to at least approximately 60 keV.
The calibration function is expressed as a combination of a non-linear component and a linear component \cite{JAKUBEK2011S262,Turecek_2013}.
We assume that the linear component extends beyond 60 keV.

A water-cooled copper heat sink was employed as the cooling system to prevent noise variations caused by temperature changes of the detectors as well as to maintain consistent thermal diffusion of the charge cloud generated by incident charged particles \cite{Chargeshare2}.
The detector temperature was maintained between 18 and 20 $^\circ$C by adjusting the temperature setting of the water-cooled chiller, which circulated water.\\
\indent
We used manipulators capable of adjusting the detector's position with an accuracy of 2 $\mathrm{\othermu m}$.
The uncertainty of incident proton's energy resulting from this accuracy of the manipulators is estimated to be below 0.1 $\mathrm{keV}$ \cite{SRIM} across the entire energy range.

\subsection{Results}
\begin{figure}[h!]
    \centering
    \hspace{-23pt}
    \begin{subfigure}{1.0\linewidth}
	\includegraphics[width=\linewidth]{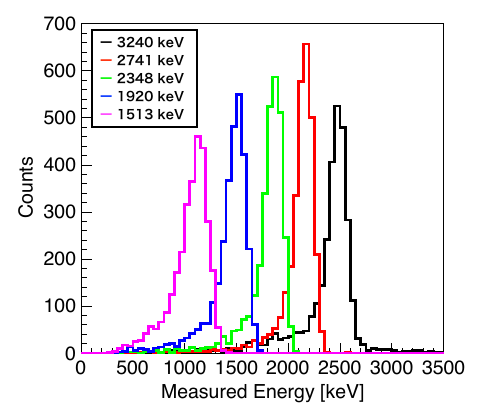}
    \end{subfigure}
    \vspace{-5pt}
    \caption{Energy spectrum of protons in the range of 1.5 to 3.3 $\mathrm{MeV}$. The legend indicates the expected energy of incident protons.}
     \vspace{10pt}
    \label{energyspectrum}
\end{figure}

\begin{figure}[h!]
    \centering
    \begin{subfigure}{1.0\linewidth}
	\includegraphics[trim={150pt 60pt 130pt 50pt},clip,width=\linewidth]{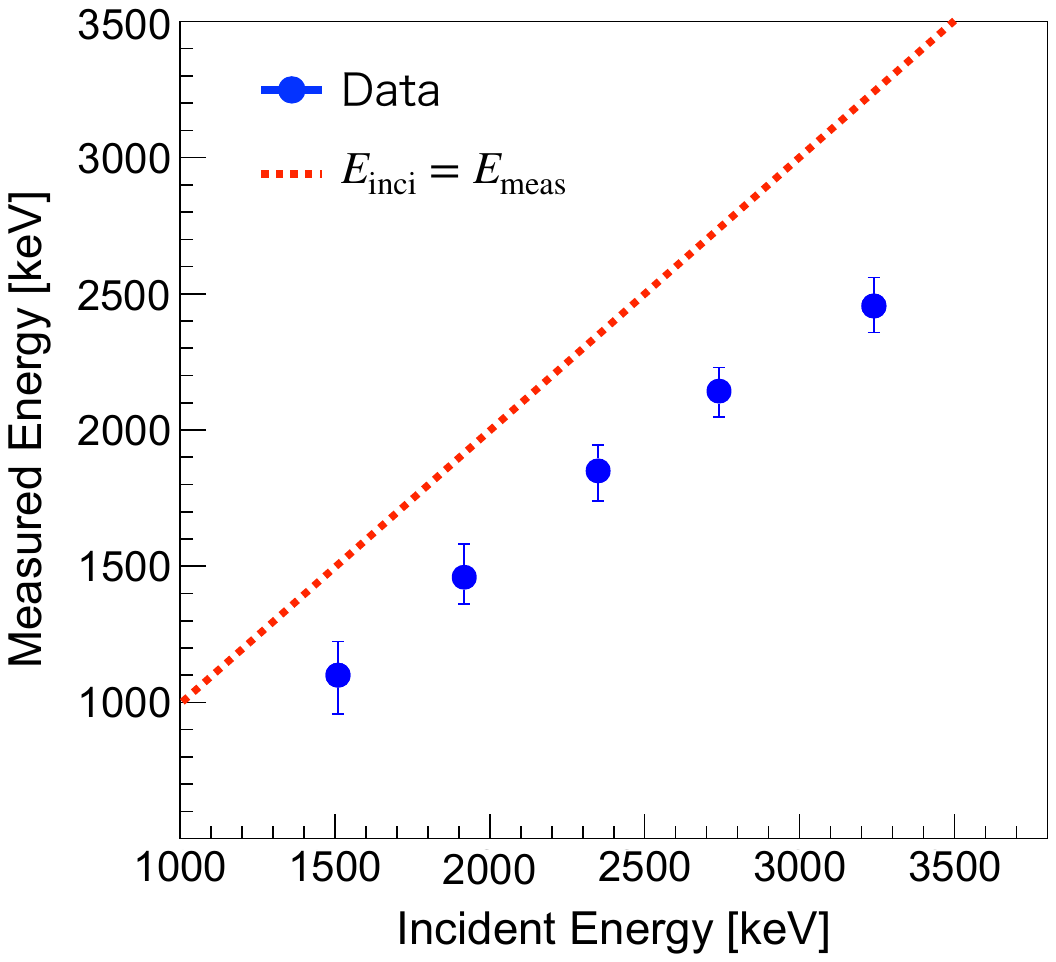}
    \end{subfigure}
    \caption{The peak position of the measured energy corresponding to the incident energy of proton in the range of 1.5 to 3.3 MeV. The blue plot shows the data from the experiment
and The red dashed line represents the relationship where the measured energy equals the incident energy.}
    \label{peakposition energyspectrum}
\end{figure}

\begin{figure}[h!]
    \centering
    \begin{subfigure}{0.48\linewidth}
	\includegraphics[width=\linewidth]{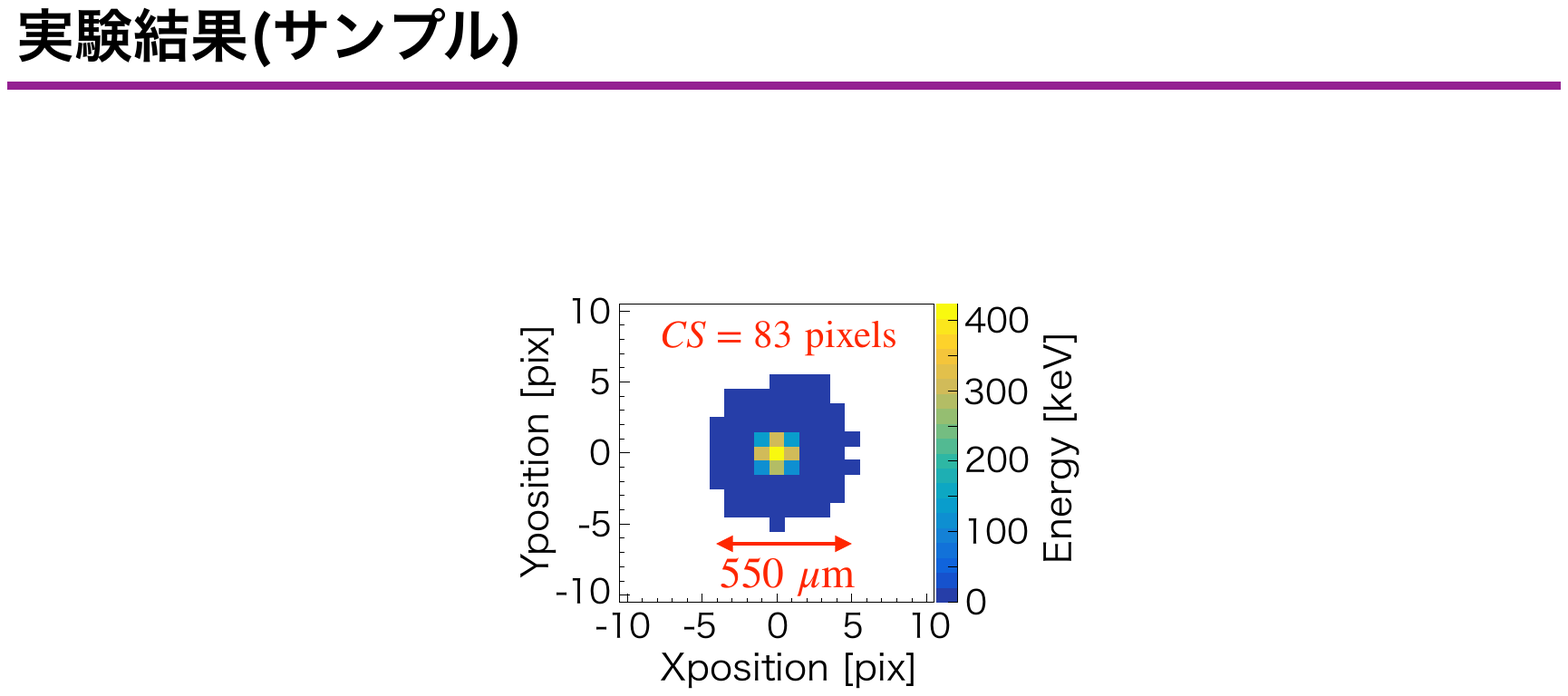}
        \caption{}
        \label{hitimage_sample_3240keV}
    \end{subfigure}
    \centering
    \begin{subfigure}{0.48\linewidth}
	\includegraphics[width=\linewidth]{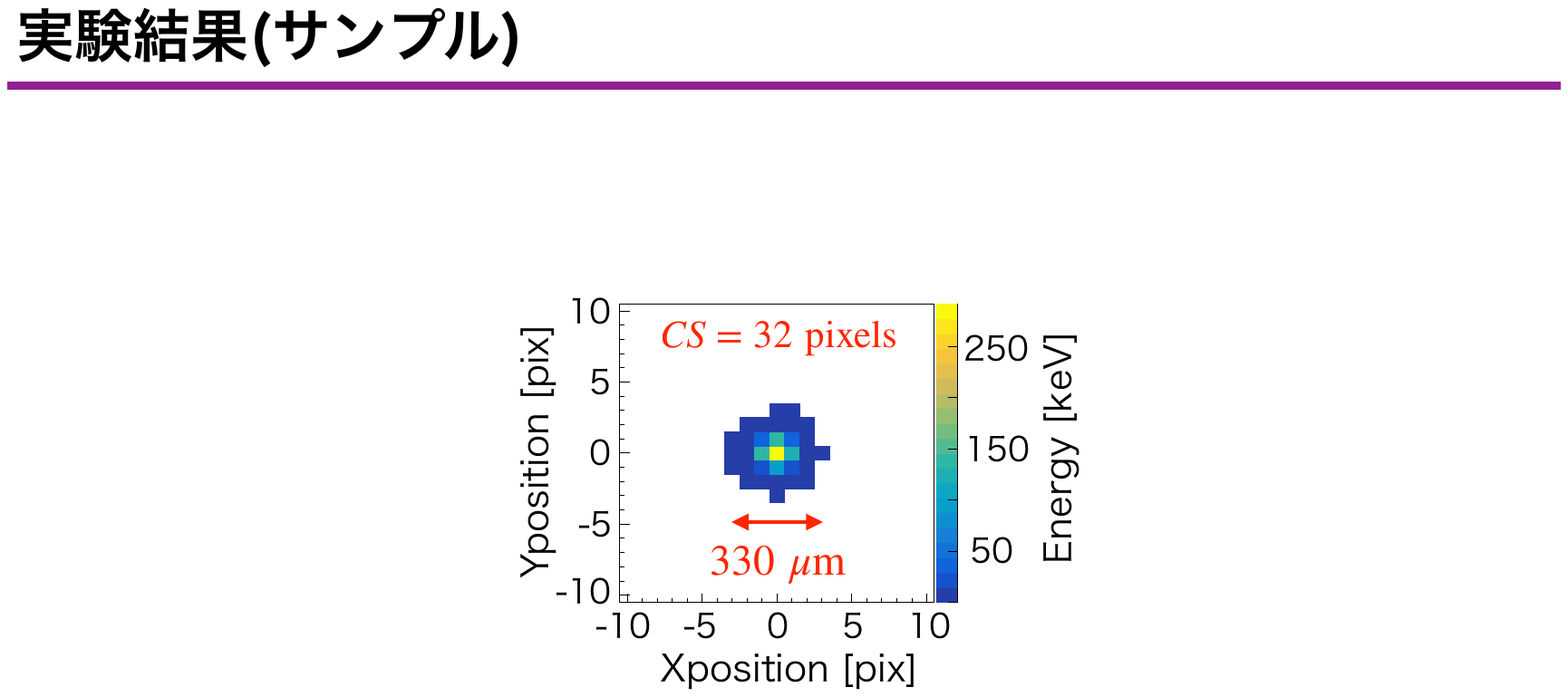}
        \caption{}
        \label{hitimage_sample_1513keV}
    \end{subfigure}
    \caption{Sample cluster images resulting from the measurement of a proton of (a) 3240 $\mathrm{keV}$ and (b) 1513 $\mathrm{keV}$ ($CS=$ cluster size).}
    \label{image}
\end{figure}

\begin{figure}[h!]
    \centering
    \includegraphics[width=\linewidth]{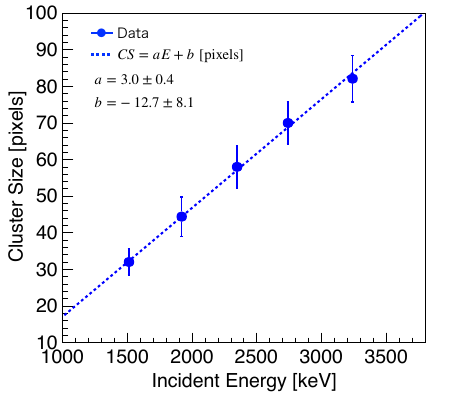}
    \caption{The cluster size corresponding to the incident energy of proton in the range of 1.5 to 3.3 $\mathrm{MeV}$. The blue plot shows the data from the experiment and the blue dash line shows the fitting result of the blue plot using a linear function.}
    \label{plot_incident_cluster}
\end{figure}

Figure \ref{energyspectrum} shows the measured energy spectrum for each incident energy.
Events that produced cluster images consisting of 9 or more pixels were treated as protons.
The measured energy values of the protons were calculated as the sum of the energy values of all pixels comprising the cluster image.
The legend in Figure \ref{energyspectrum} lists the incident energy for each spectrum.
Figure \ref{peakposition energyspectrum} indicates the measured energy depending on the incident energy.
The blue plot shows the peak position of the measured energy spectrum.
The error bar of each plot represents the FWHM (Full Width at Half Maximum) of the spectrum.
The red dashed line indicates the equality between the measured energy and the incident energy.
In all spectra, the peak position is shifted to a lower energy than the incident energy.
The shift becomes larger for higher incident energies. 
This effect occurs because higher incident energies result in larger deposited energies per pixel, increasing the likelihood of such as ToT saturation and volcano effect.
Therefore, it is difficult to evaluate the energy of charged particles around the $\mathrm{MeV}$ range from the ToT value, and a non-linear calibration function for each pixel is necessary to mitigate this issue.
The tail component on the low-energy side observed in all spectra is due to protons that scattered or passed through areas other than the capillary lid, resulting in significant energy loss.\\
\indent
Figure \ref{hitimage_sample_3240keV} and \ref{hitimage_sample_1513keV} respectively show typical cluster images for protons of 3420 $\mathrm{keV}$ and 1513 $\mathrm{keV}$.
Each cluster image consists of 83 and 32 pixels, respectively.
The shape of the cluster images is close to a circle.
The image spread range was approximately 550 $\mathrm{\othermu m}$ for a proton of 3420 $\mathrm{keV}$ and 330 $\mathrm{\othermu m}$ for a proton of 1513 $\mathrm{keV}$.
Figure \ref{plot_incident_cluster} shows the cluster size depending on the incident energy.
The blue plot shows the mean value of cluster size's distributions.
The error bar of each plot indicates the FWHM of the cluster size's distribution.
The linear fit applied to the plot yields a relationship
of the form $CS=aE+b$ with $a=(3.0\pm0.4)\times 10^{-2},~b=-12.7\pm8.1$, where CS represents the cluster size and $E$ represents the incident energy of the charged particles in $\mathrm{keV}$.
The proportionality between the cluster size and the incident energy suggests that the cluster size can be used as an alternative energy evaluation method to the ToT measurement.

\subsection{Comparison of the experimental cluster size and diffusion of charge clouds owing to the random thermal motion}
We compared the size of the cluster images in the experiment with the diffusion of the charge clouds caused by the random thermal motion.
The diffusion of charge clouds is assumed to follow a Gaussian distribution with a standard deviation $\sigma$ calculated as follows: 
\begin{equation}
    \hspace{70pt}
    \sigma = \sqrt{2Dt},
\end{equation}
where $D$ is the thermal diffusion constant of the charge carriers and $t$ is the drift time.
$D$ is calculated using Einstein's relation as follows:
\begin{equation}
    \hspace{70pt}
    D = \frac{kT}{q}\mu,
\end{equation}
where $k$ is the Boltzmann constant, $T$ is the temperature of the silicon sensor, $q$ is the elementary charge, and $\mu$ \cite{carrier_transportation} is the mobility of the charge carriers.\\
\indent
The drift time evaluation of the charge carriers requires to estimate the electric field strength $E$. 
$E$ is defined as follow:
\begin{equation}\label{e=vd}
    \hspace{70pt}
    E = \frac{V}{d},
\end{equation}
where $V$ is the reverse voltage, and $d$ is the thickness of sensor.
We assumed that $E$ is constant in this calculation.
The silicon sensor in this TimePix3 detector becomes fully depleted at a reverse voltage of 165 V.
In this situation, the fixed charges due to the impurities form an internal electric field, in addition to the electric field caused by the reverse voltage. Consequently, the electric field throughout the sensor is not uniform.
However, based on the previous work, the average electric field inside sensor is equivalent to the electric field induced by the reverse voltage \cite{3Dtrack}.
Additionally, there is no difference in the maximum carrier drift time whether the internal electric field is considered or not.
Therefore, this assumption does not give a significant impact to the estimation of the diffusion area considering the thermal random motion.
The velocity of the charge carriers is obtained from the mobility $\mu$ as follows:
\begin{equation}\label{v=µE}
    \hspace{70pt}
    v = \mu E,
\end{equation}
Using Equations \ref{e=vd} and \ref{v=µE}, the drift time $t$ is derived as follows:
\begin{equation}\label{drift time}
    \hspace{70pt}
    t = \frac{d}{v}=\frac{d}{\mu E}.
\end{equation}

\indent
The physical parameters $k$, $q$, and $d$ are $1.38 \times 10^{-23}~\mathrm{m^2~kg~s^{-2}~K^{-1}}$, $1.6 \times 10^{-19}~\mathrm{C}$, and 500 $\mathrm{\othermu m}$, respectively.
For other parameters, including the diffusion constant for the random thermal motion of the electrons/holes $D_{\mathrm{e}}$/$D_{\mathrm{h}}$, the mobility of the electrons/holes $\mu_{\mathrm{e}}$/$\mu_{\mathrm{h}}$, $T$, and $V$, the values, as listed in Table \ref{physical parameters} in Section \ref{Section3}, were used.\\
\indent
Consequently, when considering the proton incidence of 3.3 $\mathrm{MeV}$, -- the range of the trajectory is about 110 $\mathrm{\othermu m}$ from the common electrode --, the maximum drift time of the holes is estimated to be around 31 $\mathrm{ns}$ and the charge cloud extent at the end of the charge carrier drift $\sigma_{\mathrm{h}}$ is approximately 9 $\mathrm{\othermu m}$.
On the other hands, the maximum drift time of the electrons is estimated to be around 2.5 $\mathrm{ns}$ and the charge cloud extent at the end of the charge carrier drift $\sigma_{\mathrm{e}}$ is approximately 4 $\mathrm{\othermu m}$.
We found that the cluster size in the experiment is much larger than the one estimated from the diffusion due to random thermal motion.
In Section \ref{Section3}, we will estimate the diffusion of charge clouds, accounting for thermal random motion and Coulomb repulsion, using a charge drift simulation and compared these results with the experimental data to understand the physical process behind the cluster image formation.

\section{Estimation of the Cluster Size using Simulation}\label{Section3}


\subsection{Simulation setup}
We used SSD (SolidStateDetectors.jl) \cite{Abt:2021SSD}, an open-source Julia package, to simulate the charge carrier transport in a solid state detector together with the corresponding induced charge on each electrode.
Physical quantities in the solid state detector are calculated in 2D or 3D using an SOR (Successive Over Relaxation) method computed over an adaptive grid.\\
\indent
The simulation procedure can be divided into three main parts \cite{Abt:2021SSD}:
1) The construction of the detector geometry and the definition of the simulation conditions.
2) The calculation of the electric potential and electric field as well as the weighting potential of each electrode for the defined geometry. 
3) The simulation of the charge carrier drift and the calculation of the induced charge on each electrode.

\subsubsection{Geometry and physical parameters}\label{geometry_parameter}
\begin{figure}[h!]
    \centering
    \includegraphics[width=\linewidth]{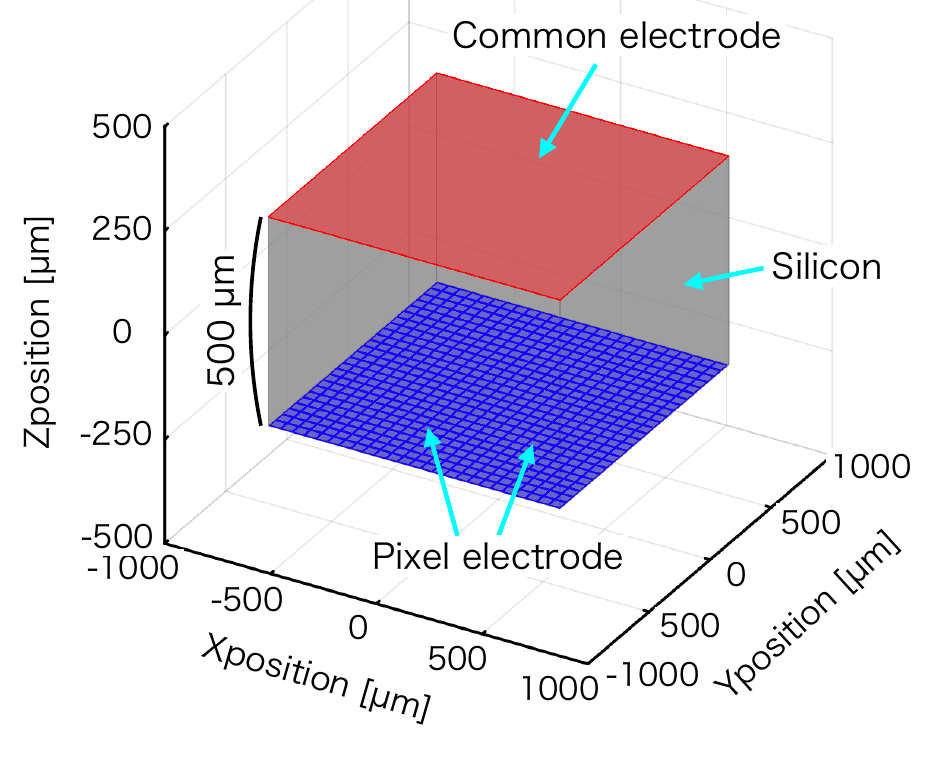}
    \caption{Geometry assuming the TimePix3 detector with a 500$~\mathrm{\othermu m}$-thick silicon-based sensitive layer.}
    \label{geometry}
\end{figure}

\begin{figure}[h!]
    \centering
    \begin{subfigure}{1.0\linewidth}
	\includegraphics[width=\linewidth]{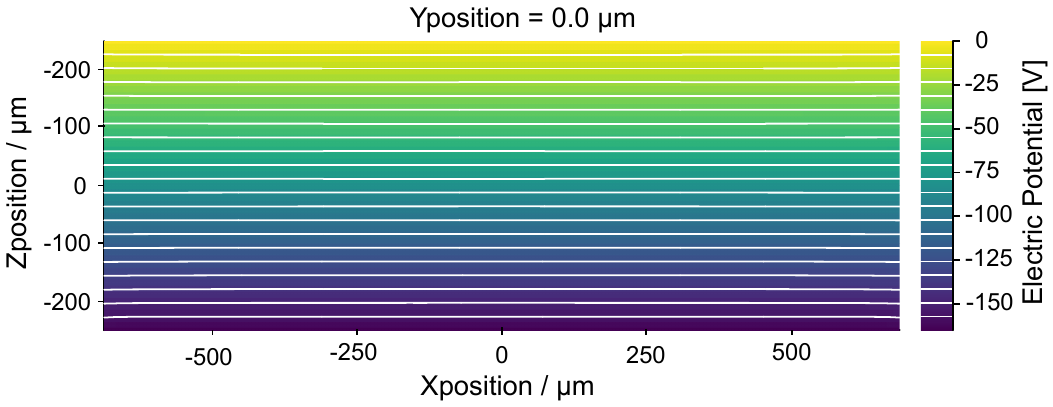}
        \caption{}
        \label{electric field}
    \end{subfigure}

    \vspace{8pt}
    
    \centering
    \begin{subfigure}{1.0\linewidth}
	\includegraphics[width=\linewidth]{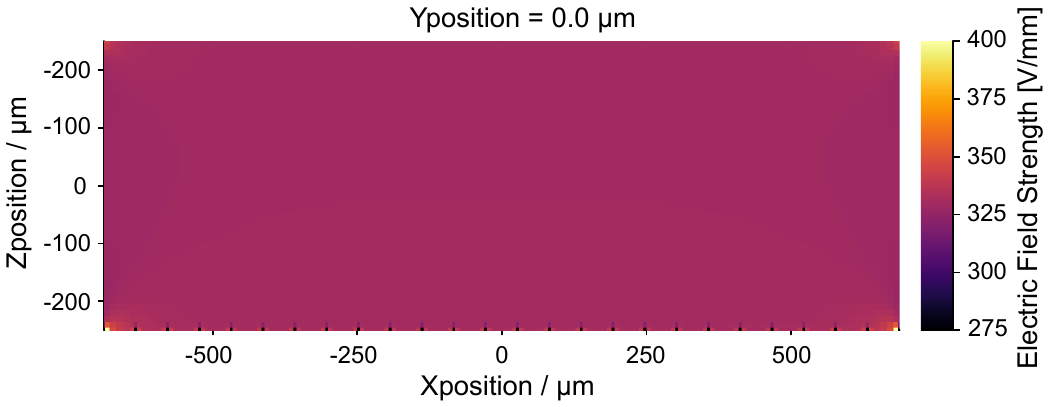}
        \caption{}
        \label{electric potential}
    \end{subfigure}

    \vspace{8pt}
    
    \centering
    \begin{subfigure}{1.0\linewidth}
	\includegraphics[width=\linewidth]{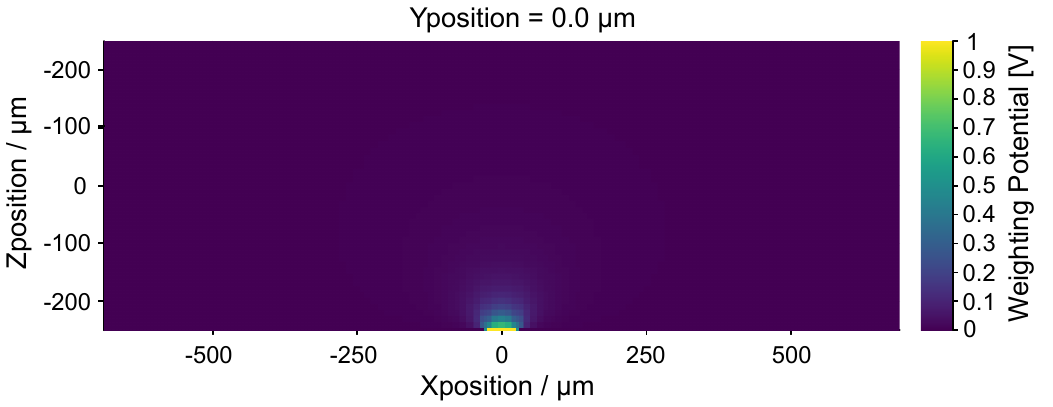}
        \caption{}
        \label{weighting potential1}
    \end{subfigure}
    \caption{(a) Electric potential, (b) electric field and (c) weighting potential of the center pixel ($(x, y)=(0,0)$) in the plane defined by Yposition = 0.0 $\mathrm{\othermu m}$. White lines in (b) represent equipotential lines.}
    \label{abcd}
\end{figure}

Figure \ref{geometry} shows the geometry assuming the TimePix3 detector with a 500 $\mathrm{\mu m}$ silicon sensor.
The geometry consisted of a 500 $\mathrm{\othermu m}$ silicon bulk (gray box), a common electrode (red plate), pixel electrodes (blue plate).
The volumes and materials associated to each geometry component are summarized in Table \ref{components of geometry}.
The TimePix3 detector used in the experiment is comprised of a n-type silicon bulk together with a p-type silicon implant divided into 256 $\times$ 256 pixels \cite{volcano1}.
However, a simplified structure with a 500 $\mathrm{\othermu m}$ silicon bulk sensor and 25 $\times$ 25 pixels was used in the simulation.
The number of pixels are decreased to reduce the calculation time for the electric potential, the electric field, and the weighting potential.
From the experimental results, the cluster image measured in the experiment extends to approximately 10 $\times$ 10 pixels at most; thus, the geometry with 25 $\times$ 25 pixels is sufficiently large for estimating the cluster size.\\
\indent
Moreover, we set several physical parameters related to the charge carrier drifts to reproduce the conditions of operation of the TimePix3 detector. 
These parameters are summarized in the Table \ref{physical parameters} and include the mobility of the electrons/holes $\mu_\mathrm{e}$/$\mu_\mathrm{h}$ \cite{carrier_transportation}, the diffusion constant for the random thermal motion of the electrons/holes $D_{\mathrm{e}}$/$D_{\mathrm{h}}$ calculated by Einstein's equation, the electron-hole pair creation energy $E_{\mathrm{ion}}$ \cite{electron-hole_pair_creation_energy}, the Fano factor $f_{\mathrm{fano}}$ \cite{fanofactor}, the supply voltage between the common electrode and the pixel electrodes, and the temperature of the silicon bulk $T$.\\
\indent
Figure \ref{geometry} shows the geometry schematic diagram.
The electric field and the electric potential of the geometry are shown in Figures \ref{electric field} and \ref{electric potential}, respectively.
Figure \ref{weighting potential1} shows the weighting potential of the center pixel, which is located at (x,y)=(0, 0) in the geometry.
The weighting potential $\phi^w{\bf (r)}$ is calculated by solving Laplace's equation $\Delta \phi^w{\bf (r)}=0$ with boundary conditions of $\phi^w=1$ at the electrode of interest and $\phi^w=0$ at all other electrodes.
In the calculation of the potential and the weighting potential, we set 1.65 V and 0.1 V as the maximum relative allowed differences of the potential value of neighbored grid points in each dimension. And we set the absolute maximum difference of 0.1 mV between two iterations across all grid points. As a result, the average grid size was approximately 10 $\mathrm{\othermu m}$ in this geometry.

\begin{table}[h!]
\centering
\caption{Dimensions and material for each component used in the geometry.}
\vspace{-8pt}
\renewcommand{\arraystretch}{1.4} 
\setlength{\doublerulesep}{1pt} 
\begin{tabular}{l l l} 
    \hline
    Component & Dimensions  [$\mathrm{\othermu m}^3$] & Material \\ 
    \hline\hline
    Silicon bulk & $1380\times1380\times500$ & Si\\ 
    Common electrode & $1380\times1380\times0.5$ & Al\\ 
    
    Pixel electrode & $50\times50\times0.5$ & Pt\\ 
    \hline
\end{tabular}
\label{components of geometry}
\vspace{3pt}
\begin{minipage}{\textwidth}
\hspace{5pt}
\footnotesize\textit{Note: The gap between each pixel electrode is 5 $\mathrm{\othermu m}$.}
\end{minipage}

\end{table}

\begin{table}[h!]
\centering
\caption{Physical parameters involved in the charge carrier drift.}
\vspace{-8pt}
\renewcommand{\arraystretch}{1.4} 
\setlength{\tabcolsep}{3pt} 
\setlength{\doublerulesep}{1pt} 
\begin{tabular}{l@{\hspace{10pt}} p{4cm}@{\hspace{-15pt}} l}  
    \hline
    Parameter & Description & Value \\ 
    \hline\hline
    $\mu_{\mathrm{e}}$ & Mobility of electrons & 1360 $\mathrm{cm^2/V \cdot s}$ \cite{carrier_transportation}\\ 
    $\mu_{\mathrm{h}}$ & Mobility of holes & 495 $\mathrm{cm^2/V \cdot s}$ \cite{carrier_transportation}\\ 
    $D_{\mathrm{e}}$ & Constant for thermal\newline diffusion of electrons & 34.3 $\mathrm{cm^2/s}$\\  
    $D_{\mathrm{h}}$ & Constant for thermal\newline diffusion of holes & 12.5 $\mathrm{cm^2/s}$\\
    $E_{\mathrm{ion}}$ & Electron-hole pair\newline creation energy & 3.62 $\mathrm{eV}$ \cite{electron-hole_pair_creation_energy} \\
    $F_{\mathrm{fano}}$ & Fano factor & 0.1 \cite{fanofactor}\\
    $V$ & Supply voltage & 165 $\mathrm{V}$\\
    $T$ & Temperature & 293 $\mathrm{K}$\\
    \hline
\end{tabular}
\label{physical parameters}
\end{table}

  \subsubsection{Initial distribution of charge carriers}
Before the start of the charge carrier drift simulation, the initial distribution of the charge carriers created by the incident charged particles must be determined.
The trajectory of the charged particles in the sensor was considered to be straight and perpendicular to the detector's surface in the simulation, although it is bent by interactions with nuclei and electrons of the sensitive layer in reality.
The initial position of the charge clouds was set above the center pixel.
To determine the initial distribution of the charge carriers, SRIM was employed \cite{SRIM}. 
The number of charge carriers generated at each position was determined by calculating the energy deposited at each position using SRIM and then dividing it by the electron-hole pair creation energy $E_{\mathrm{ion}}$ \cite{electron-hole_pair_creation_energy}.
Although the actual charge cloud created in the detector by the incident charged particles is continuous along the charged particle trajectory \cite{SOLC202279}, the initial position of the charge cloud was discretely determined according to the following two steps:
1) calculate the total number of the charge carriers generated at each 10 $\mathrm{\othermu m}$ interval along the trajectory.
2) charge clouds corresponding to the sum of the charge carriers are placed at each 10 $\mathrm{\othermu m}$ step.
The shape of each charge cloud is a sphere with a radius of 1 $\mathrm{\othermu m}$.

\subsubsection{Charge carrier drift and induced charge}
Electrons and holes drift toward their respective electrodes along the electric field calculated in Section \ref{geometry_parameter} \cite{Abt:2021SSD}. In this simulation, the random thermal motion and Coulomb self-repulsion were included as the diffusion factors of the charge carriers.
The position of each charge carrier was tracked in steps of 0.1 $\mathrm{ns}$.\\
\indent
The time evolution of the induced charge at each electrode is calculated using the weighting potential determined in Section \ref{geometry_parameter} and the Shockley-Ramo theorem \cite{Shokley-Romo1,Shokley-Romo2,Shokley-Romo3}.
In this theorem, when a charge carrier $q(\bf r)$ moves from a position $\bf r_{\mathrm{i}}$ to a new position $\bf r_{\mathrm{j}}$, the induced charge $Q$ at a given electrode is expressed as follows:
\begin{equation}
    \hspace{50pt}
    Q = -\int_{\bf r_{\mathrm{j}}}^{\bf r_{\mathrm{i}}} q({\bf r})\nabla\phi^w({\bf r}) \cdot d{\bf r},
\end{equation}
where $\phi^w({\bf r})$ is the corresponding weighting potential.

\subsection{Estimation of cluster size considering charge cloud diffusion}

\subsubsection{Method}\label{method1}
\begin{figure}[h!]
	\centering 
	\includegraphics[width=\linewidth]{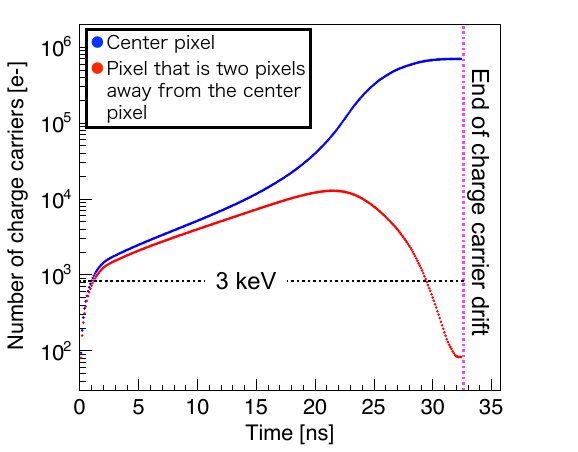}
        \caption{Time variation of the induced charge at the pixel electrodes. The blue plot shows the time variation of the induced charge on the pixel electrode that collects the charge carriers, with a final signal above 3 $\mathrm{keV}$ (center pixel located at $(x, y)=(0, 0)$). The red plot shows the time variation of the induced charge on the pixel electrode that does not collect the charge carriers, with a final signal below 3 $\mathrm{keV}$(pixel located two pixels away from the center pixel).}
        \label{waveform}
\end{figure}

First, we estimated the cluster size directly from the simulation model, including the physical process of charge cloud diffusion.
Figure \ref{waveform} shows the time variation of the induced charge at two different pixel electrodes for a 3200 $\mathrm{keV}$ proton incidence.
The blue plot in Figure \ref{waveform} shows the time variation of the induced charge on the center pixel electrode located at $(x,y)=(0,0)$ in the geometry in Figure \ref{geometry}.
The induced charge increases as the charge carrier drift proceeds, and the induced charge is above 3.0 $\mathrm{keV}$ - the threshold level of each pixel in the TimePix3 detector - at the end of all charge carrier drift.
The induced charge at the end of the charge carrier drift represents the number of charge carriers collected by the pixel electrode.
We included the pixel electrodes for which the induced charge at the end of the charge carrier drift exceeds 3.0 $\mathrm{keV}$ in the cluster image.
Moreover, the red plot shows the time variation of the induced charge on the pixel electrode located two pixels away from the center pixel in the $x$ direction.
The induced charge increases in the early stage of the charge carrier drift, but decreases in the later stage.
At the end of all charge carrier drift, the induced charge decreases to below 3.0 $\mathrm{keV}$.\\
\indent
Charge carrier drift simulations were performed from 1.4 $\mathrm{MeV}$ to 3.4 $\mathrm{MeV}$ in 100 $\mathrm{keV}$ steps, with 100 simulations per energy step.
The incident position of the proton for each trial was randomly set within the area occupied by the center pixel of the geometry in Figure \ref{geometry}.
The cluster size at each energy was calculated as the average of the cluster sizes estimated from each trial.\\
\indent

\subsubsection{Results}\label{results1}
\begin{figure}[h!]
    \centering
    \begin{subfigure}{0.48\linewidth}
	\includegraphics[width=\linewidth]{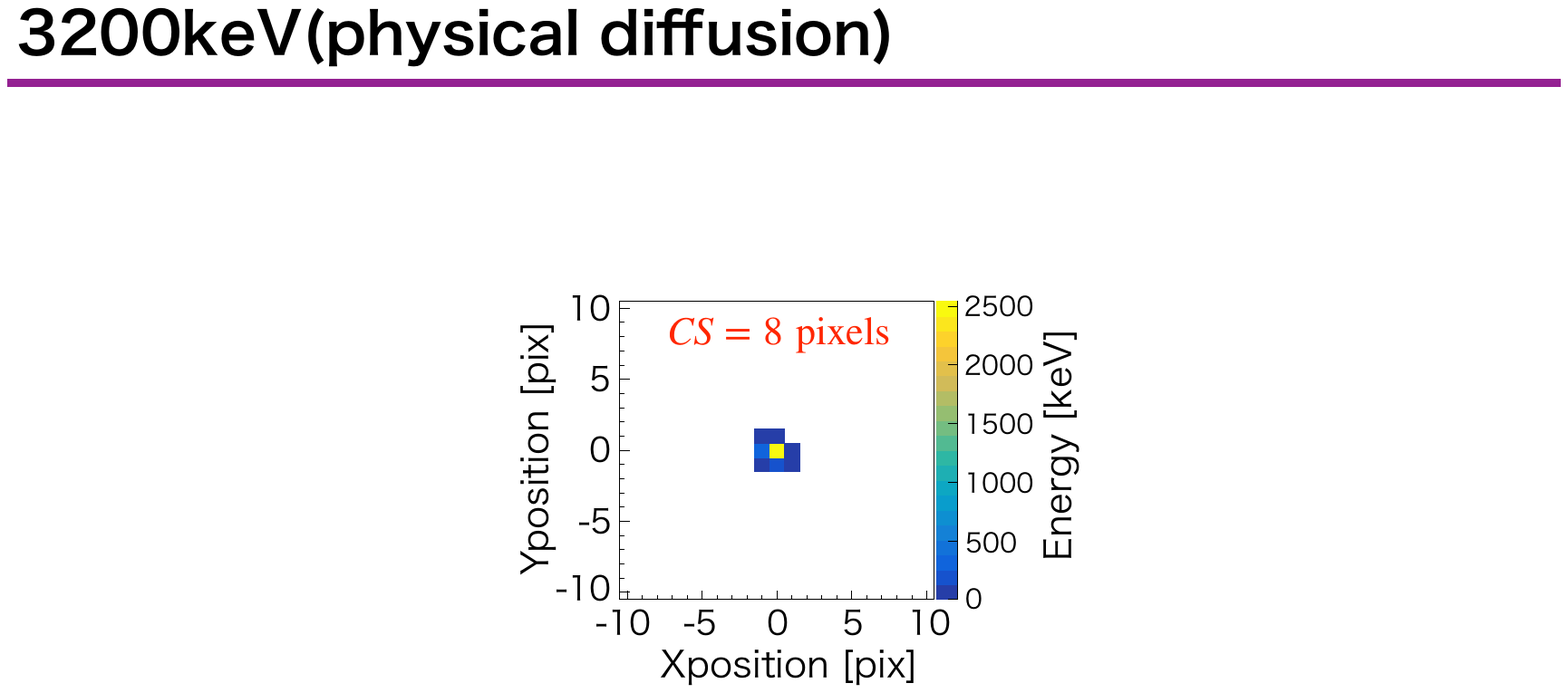}
        \caption{}
        \label{clusterimage_sim_3200keV}
    \end{subfigure}
    \centering
    \begin{subfigure}{0.48\linewidth}
	\includegraphics[width=\linewidth]{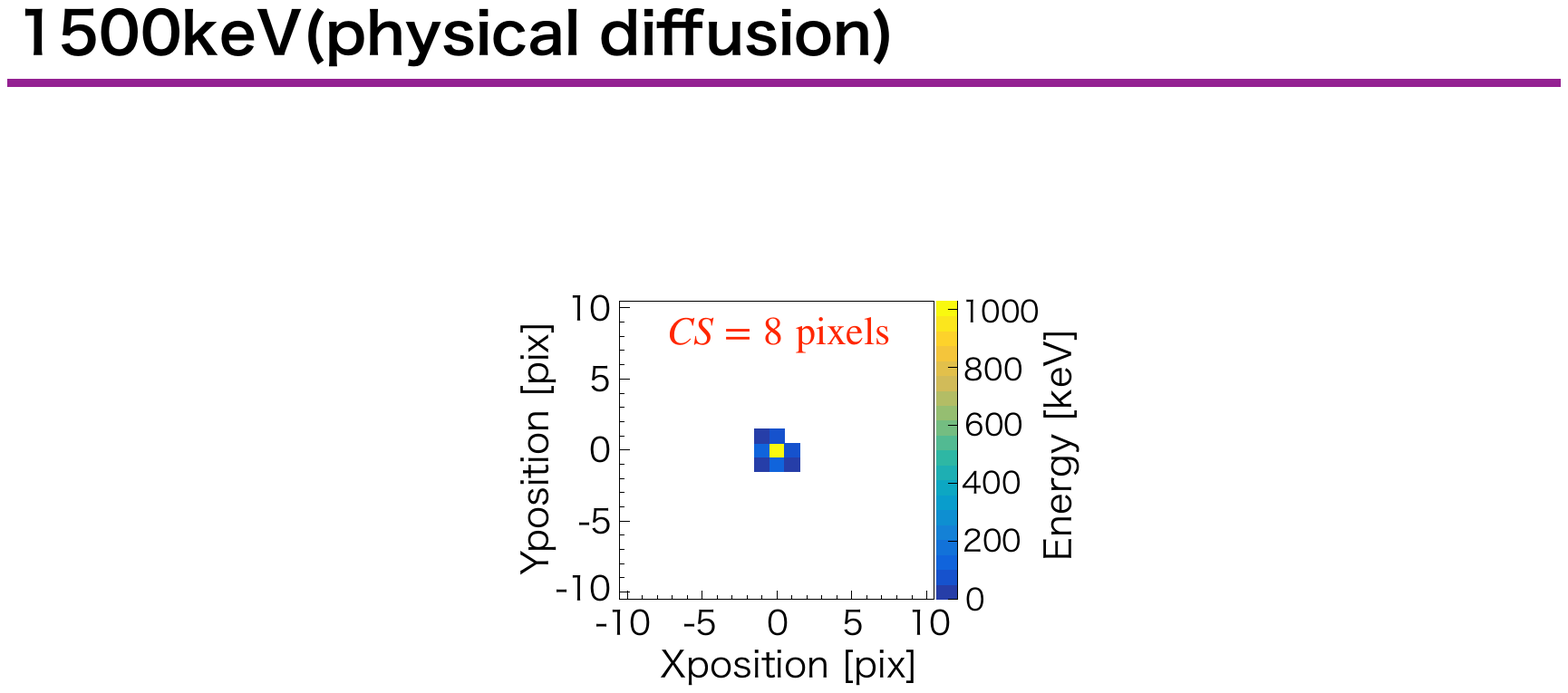}
        \caption{}
        \label{clusterimage_sim_1500keV}
    \end{subfigure}
    \caption{Estimated cluster images with the charge carrier diffusion considered. Only pixels that collected charge carriers corresponding to 3 $\mathrm{keV}$ or more are shown. a) Cluster image of a proton of 3200 $\mathrm{keV}$. b) Cluster image of a proton of 1500 $\mathrm{keV}$ ($CS=$ cluster size).}
    \label{image}
\end{figure}

Figures \ref{clusterimage_sim_3200keV} and \ref{clusterimage_sim_1500keV} show the simulated cluster images for protons with incident energies of 3200 $\mathrm{keV}$ and 1500 $\mathrm{keV}$, respectively, for one trial in which the incident position was $(x,y)=(-4.8, -1.5)$ and $(-1.4, -1.3)$, respectively.
In both figures, the energy deposition is concentrated in the center pixel and the cluster image consists of 8 pixels. \\
\indent
The estimated cluster size was found to vary between 4 and 9 pixels for energies comprised between 1.4 and 3.4 $\mathrm{MeV}$ as shown in Figure \ref{plot_incident_cluster_sim} (green plots).
These results indicate that the number of pixel electrodes that measures an energy deposition above 3.0 $\mathrm{keV}$ is comprised between 4 and 9 pixels.
The simulated cluster sizes are smaller when compared to the results of the experiment and do not increase in proportion to the energy of the protons throughout the energy range.

\subsection{Cluster size estimation considering the magnitude of transient induced signal during charge carriers collection}

\subsubsection{Method}
We concluded from the findings outlined in Section \ref{results1} that the physical diffusion of charge clouds does not explain by itself the experimental cluster size measured across the energy range.
Therefore, instead of selecting the pixels based on the induced charge at the end of charge carriers collection, we selected the pixels based on the induced charge throughout the drift to account for the induction of transient signals.\\
\indent
We prepared two quantities and determined a new condition for selecting the triggered pixel.
These two quantities are the threshold for the induced charge $Q_\mathrm{TH}$ and the time of the pulse height evaluation as $t_\mathrm{eval}$.
The determination of this new condition was conducted as follows:
1) we evaluated the induced charge at $t_\mathrm{eval}$ at each pixel.
2) If the induced charge at $t_\mathrm{eval}$ is greater than $Q_\mathrm{TH}$, the corresponding pixel is counted as a part of the image.
3) $Q_{\mathrm{TH}}$ and $t_{\mathrm{eval}}$ are chosen to minimize the difference between the experimental and simulated cluster sizes in the range of 1.5 $\mathrm{MeV}$ to 3.3 $\mathrm{MeV}$.
4) The difference between the experimental and simulated cluster sizes is evaluated using the $\eta$ value as follows:
\begin{equation}
    \hspace{30pt}
    \eta=\sum_{i=1}^5 \frac{(CS_{\mathrm{mea}}(E_i)-CS_{\mathrm{sim}}(E_i))^2}{CS_{\mathrm{sim}}(E_i)^2},
\end{equation}
where $E_i$ indicates the incident energy of the proton.
In the beam experiment, we prepared five different energies.
To calculate $\eta$, the same energies are prepared in the simulation.
$E_1, E_2, E_3, E_4$, and $E_5$ respectively correspond to 1513, 1920, 2348, 2741, and 3240 $\mathrm{keV}$.
$CS_{\mathrm{sim}}$ and $CS_{\mathrm{mea}}$ show the estimated cluster size in the simulation and the cluster size observed in the experiment, respectively.
The induced charge and the time were scanned every 0.1 $\mathrm{keV}$ (0 $\mathrm{keV}$ $\leqq$ $Q$ $\leqq$ 20 $\mathrm{keV}$) and 0.1 $\mathrm{ns}$ (0 $\mathrm{s}$ $\leqq$ $t$ $\leqq$ 30 $\mathrm{s}$) to investigate the minimum $\eta$ value.\\
\indent
The cluster size at each energy was calculated as the average of the cluster sizes estimated across 100 trials similarly, as described in Section \ref{method1}.

\subsubsection{Results}
\begin{figure}[h!]
    \centering
    \begin{subfigure}{0.48\linewidth}
	\includegraphics[width=\linewidth]{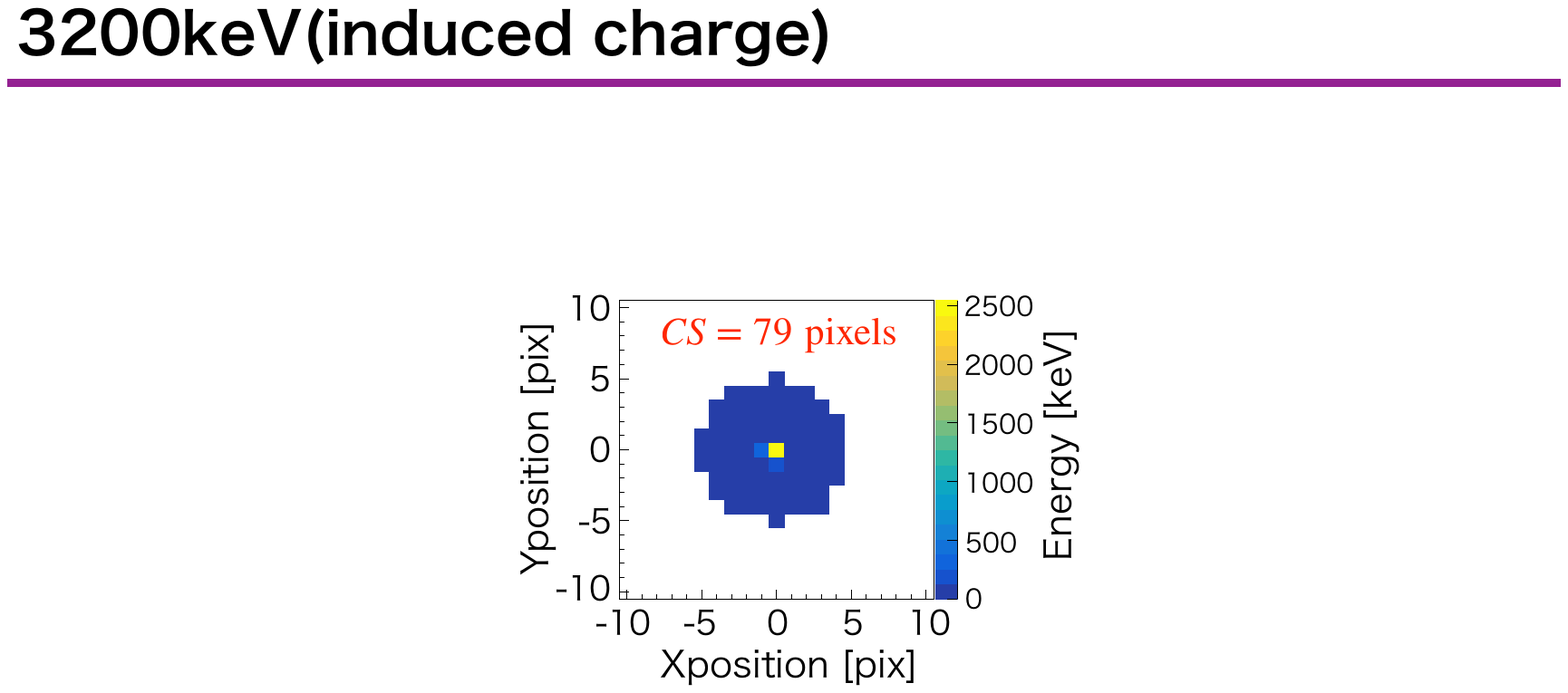}
        \caption{}
        \label{clusterimage_sim_3200keV_new_method}
    \end{subfigure}
    \centering
    \begin{subfigure}{0.48\linewidth}
	\includegraphics[width=\linewidth]{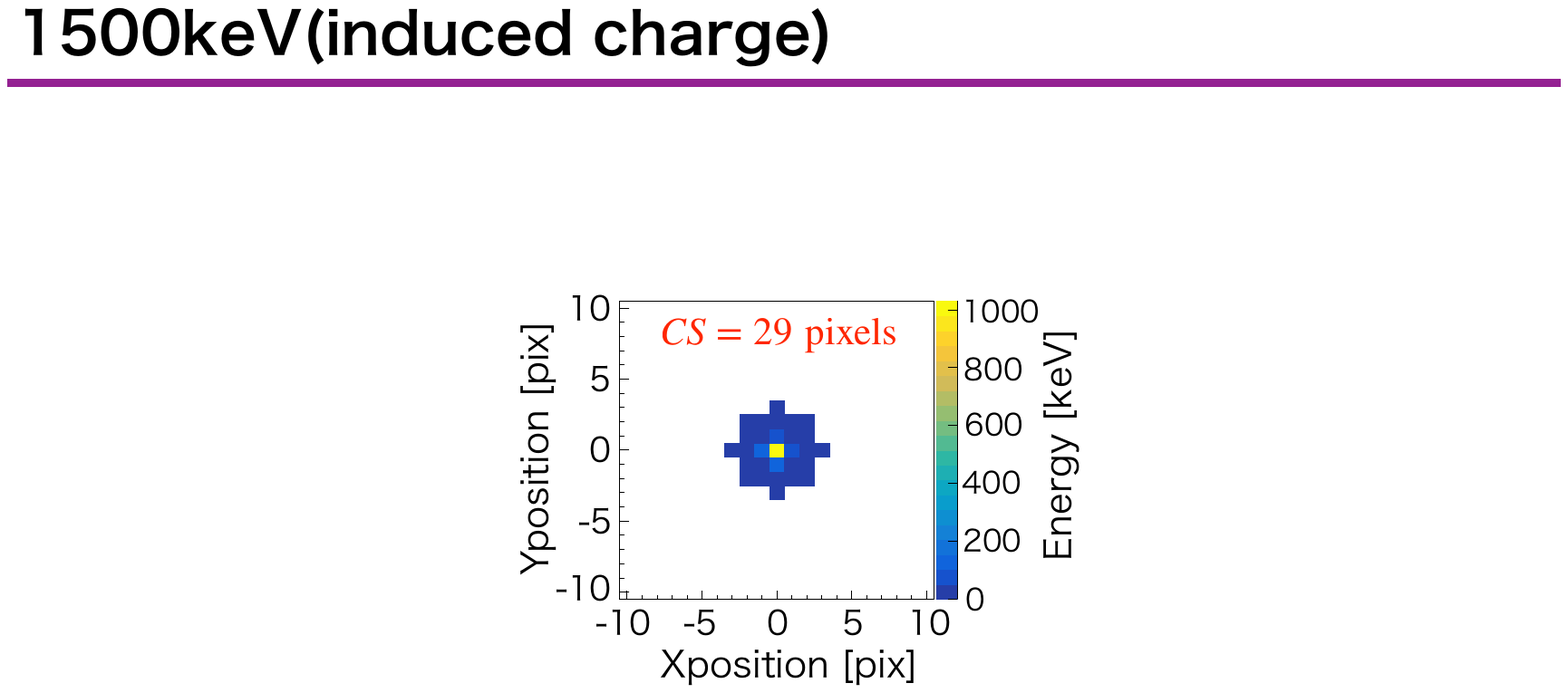}
        \caption{}
        \label{clusterimage_sim_1500keV_new_method}
    \end{subfigure}
    \caption{Estimated cluster images considering the magnitude of induced charge during the charge carriers collection. Only pixels for which the induced charge corresponding to an energy of 6.4 $\mathrm{keV}$ or more after a time of 15.8 $\mathrm{ns}$ are shown. a) Cluster image of a proton of 3200 $\mathrm{keV}$. b) Cluster image of a proton of 1500 $\mathrm{keV}$ ($CS=$ cluster size).}
    \label{clusterimage_sim_new_method}
\end{figure}

\begin{figure}[h!]
    \centering
    \includegraphics[trim={120pt 60pt 110pt 50pt},clip,width=\linewidth]{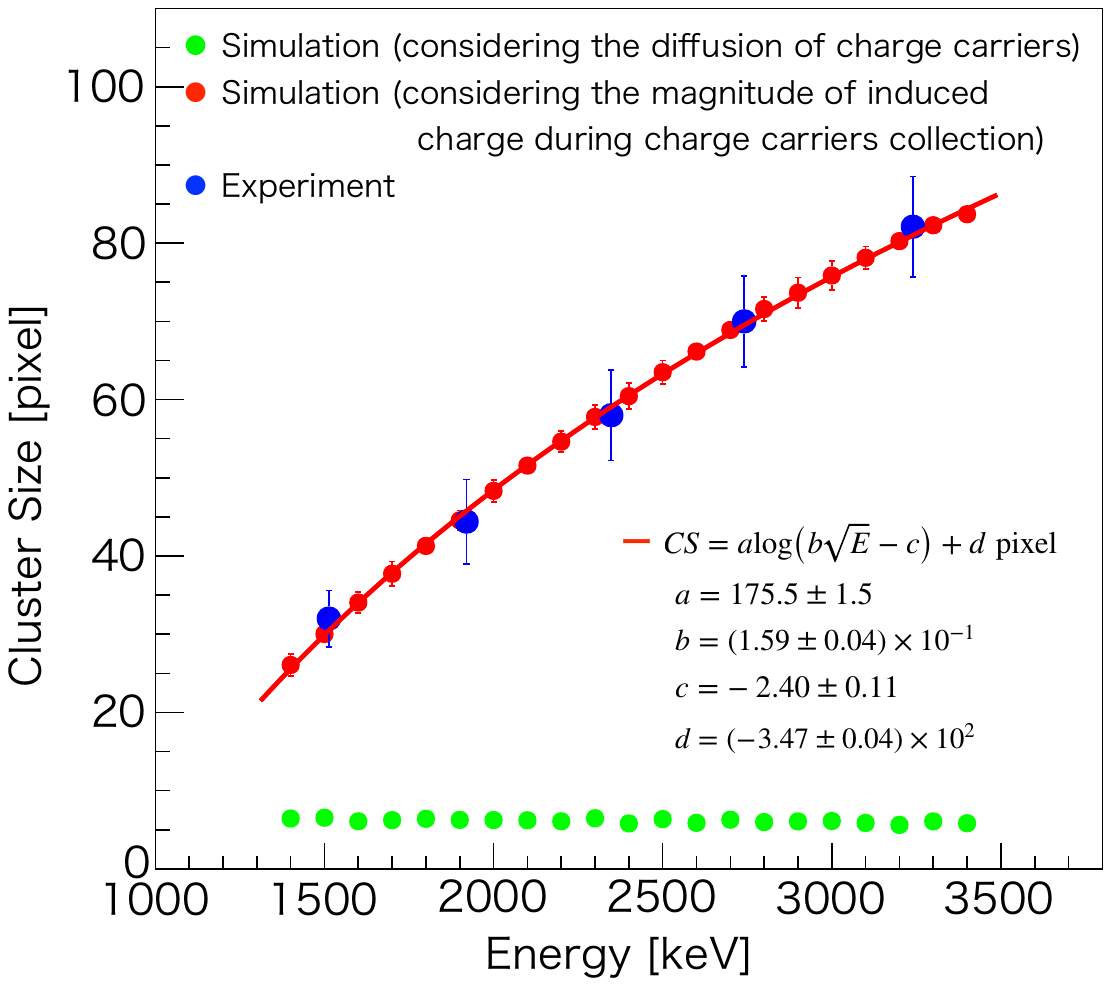}
    \caption{Cluster size measured in the experiment (blue plot) and estimated cluster size in the simulation for proton of energies comprised between 1.4 $\mathrm{MeV}$ and 3.4 $\mathrm{MeV}$. The red line represents the fitting result of the red plots. $CS$ and $E$ correspond to the cluster size and the energy of proton, respectively.}
    \label{plot_incident_cluster_sim}
\end{figure}

\begin{figure}[h!]
    \centering
    \begin{subfigure}{1.0\linewidth}
	\includegraphics[trim={120pt 60pt 110pt 50pt},clip,width=\linewidth]{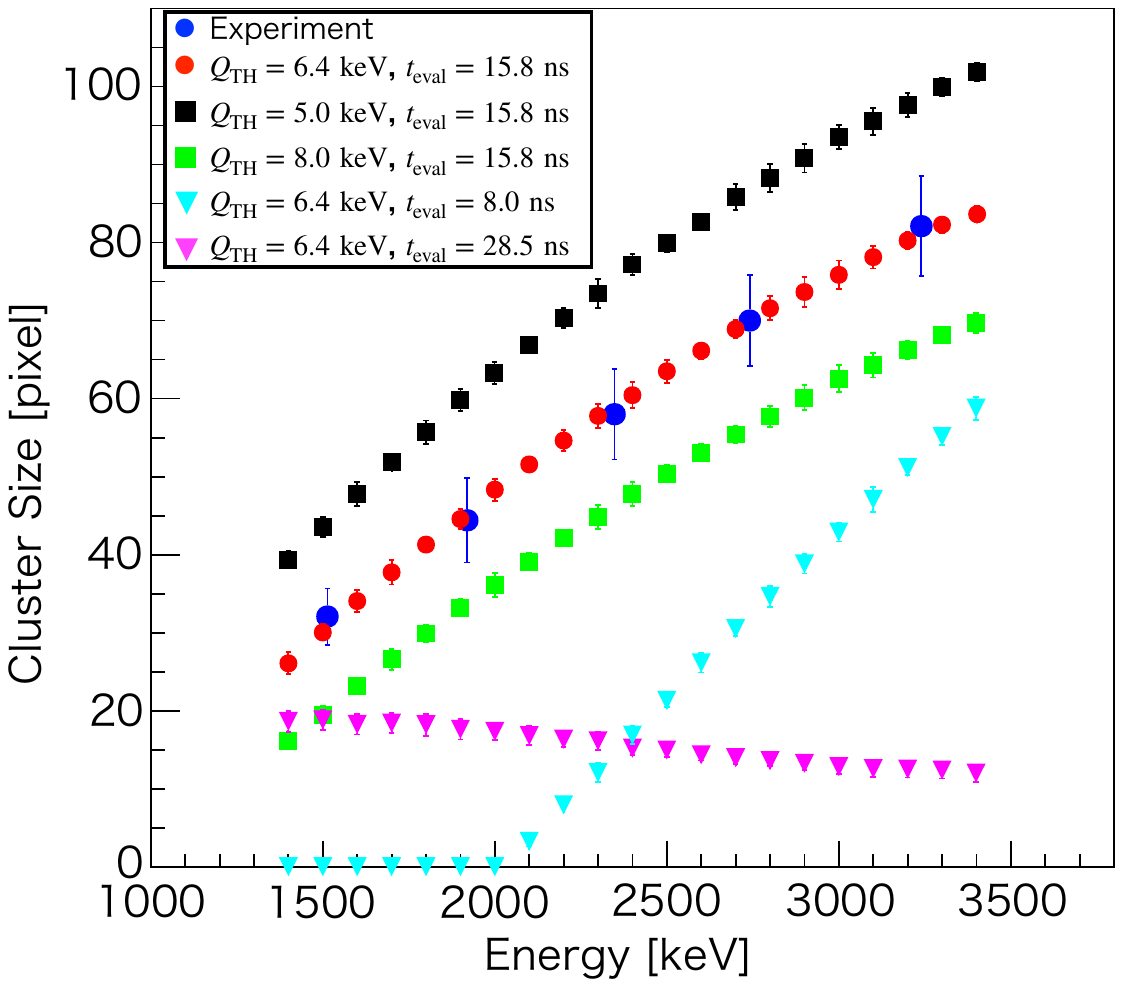}
    \end{subfigure}
    \caption{Cluster size measured in the experiment (blue plot) and Cluster size when five different combinations of $Q_\mathrm{TH}$ and $t_\mathrm{eval}$ were applied ( red plot, black plot, green plot, cyan plot, magenta plot). Five different combinations were prepared: ($Q_{\mathrm{TH}}, t_{\mathrm{eval}}$) = (6.4 keV, 15.8 ns), (5.0 keV, 15.8 ns), (8.0 keV, 15.8 ns), (6.4 keV, 8.0 ns), and (6.4 keV, 28.5 ns).}
    \label{variouspattern}
\end{figure}

The combination of $Q_{\mathrm{TH}}$ = 6.4 $\mathrm{keV}$ and $t_{\mathrm{eval}}$ = 15.8 $\mathrm{ns}$ was found to minimize the difference between the cluster size in the experiment and the estimated cluster size in the simulation.
Figures \ref{clusterimage_sim_3200keV_new_method} and \ref{clusterimage_sim_1500keV_new_method} show the cluster images when this new method was applied to the same trial data used to create Figures \ref{clusterimage_sim_3200keV} and \ref{clusterimage_sim_1500keV}, respectively.
The energy value of each pixel represents the magnitude of the induced charge at the end of the charge carriers collection.
However, for pixels with an physical energy deposition of less than 3.0 $\mathrm{keV}$, the energy value was consistently set to 3.0 $\mathrm{keV}$.
The estimated cluster size increased between 20 and 90 pixels monotonically in the range of 1.4 $\mathrm{MeV}$ to 3.4 $\mathrm{MeV}$ (red plots in Figure \ref{plot_incident_cluster_sim}).
We found that this method reproduces cluster sizes and energy dependencies similar to those of the experiment.\\
\indent
In Figure \ref{variouspattern}, we present the cluster size for different combinations of $Q_{\mathrm{TH}}$ and $t_{\mathrm{eval}}$. 
Five different combinations were prepared: ($Q_{\mathrm{TH}}, t_{\mathrm{eval}}$) = (6.4 keV, 15.8 ns), (5.0 keV, 15.8 ns), (8.0 keV, 15.8 ns), (6.4 keV, 29.0 ns), and (6.4 keV, 8.0 ns).
When $t_{\mathrm{eval}}$ is fixed and $Q_{\mathrm{TH}}$ is varied, a smaller $Q_{\mathrm{TH}}$ results in more triggered pixels, leading to a larger cluster size. Conversely, a larger $Q_{\mathrm{TH}}$ causes fewer pixels to be triggered, reducing the cluster size.
When $Q_{\mathrm{TH}}$ is fixed and $t_{\mathrm{eval}}$ is varied, a smaller $t_{\mathrm{eval}}$ results in fewer triggered pixels because the induced charge generated in each pixel is insufficient. 
Similarly, when $t_{\mathrm{eval}}$ is larger, the cluster size decreases because the induced charge in pixels collecting fewer charge carriers becomes small near the end of charge collection, preventing those pixels from being triggered.
In this case, the cluster size is smaller as the incident energy is larger.
This is because the phase of the induced charge development varies depending on the incident energy as shown. 
Protons with higher energy generate charge carriers closer to the pixel electrodes, enabling faster the charge induction.
So, in the later phase, the induced charge becomes small and many pixel are not triggered.
On the other hand, lower-energy protons generate charge carriers near the common electrode, requiring a longer time to reach the pixel electrodes and the development of the induced charge is also later.
As a result, in later phase, the induced charge still be large and some pixel are triggered.

\section{Discussion}\label{discuss}
As shown in Figure \ref{plot_incident_cluster_sim}, the cluster size estimated in the simulation was smaller than that observed in the experiment when the physical diffusion, including the thermal diffusion and the Coulomb self-repulsion, was considered as the factor in the formation of the cluster image.
Additionally, the estimated cluster size remained constant across the energy range.
However, when considering that the energy registration occurs owing to the transient induced charge at each pixel electrode, the cluster size estimated in the simulation reproduced the experimental results across the measured energy range.
The relationship between the proton energy and the cluster size obtained through this estimation method can be described in the investigated energy range using the following function:
\begin{equation}
\hspace{50pt}
CS=a~\mathrm{log}(b\sqrt{E}-c)+d,
\end{equation}
where $a=175.5\pm 1.5$, $b=(1.59\pm 0.04)\times 10^{-1}$, $c=-2.40\pm 0.11$, and $d=(-3.47\pm 0.04)\times 10^2$.
This model is in good agreement with the experimental results within the error bar.\\
\indent
These results indicate that the energy registration in the TimePix3 detector is not determined by the amount of charge carriers collected at each pixel electrode, but by the magnitude of the transient induced charge at each pixel electrode during the charge carriers collection.
The transient induced charge occurs even in pixels that do not collect the charge carriers.
These pixels contribute to the cluster image, which is larger than the extent of the physical diffusion of charge clouds.
Moreover, when the number of charge carriers in the sensor get larger, the transient induced charge increases and more pixels measure a pulse exceeding the threshold value.
Consequently, we observe an increase in cluster size proportional to the incident energy of the protons in the experiment.

\section{Conclusion}
Through the proton irradiation experiment and carrier drift simulation, we found that the cluster size of charged particles measured with the TimePix3 detector with a 500 $\mathrm{\othermu m}$ silicon sensor can be used to evaluate the energy deposited by charged particles in the detector.
Moreover, we discovered that the TimePix3 detector's ability to capture the induced charge at a pixel electrode during the charge carrier collection contributes to the energy dependency of the cluster size. 
Considering that the TimePix3 detector is widely used in various fields, including particle and nuclear physics experiments, this energy evaluation method is particularly effective for detecting high-energy heavy ions prone to cause ToT saturation and volcano effects.


\section*{Acknowledgements}
This work was supported by JSPS, Japan KAKENHI Grant Numbers 22J12583, 18H05463, 20H00153, 24H00244, 23K13130, 23H00120, and 18H05461.
S.N. has also been supported by Overseas Research Fellowship by JSPS and FoPM (WISE Program) and JSR Fellowship, the University of Tokyo, Japan.
The experiment was performed at the Pelletron facility (joint-use 
equipment) at the Wako Campus, RIKEN.
\appendix

\bibliographystyle{unsrt}
\bibliography{reference}






\end{document}